\documentclass[epj-spec]{svjour}
\usepackage{graphics}
\begin{document}

\newcommand{\bl}[1]{\begin{equation}\label{#1}}
\newcommand{\be}{\begin{equation}}
\newcommand{\ee}{\end{equation}}
\newcommand{\bea}{\begin{eqnarray}}
\newcommand{\eea}{\end{eqnarray}}

\newcommand{\pd}[2]{\frac{\partial{#1}}{\partial{#2}}}
\newcommand{\td}[2]{\frac{\mathrm{d}{#1}}{\mathrm{d}{#2}}}
\newcommand{\rec}[1]{\frac{1}{#1}}
\newcommand{\z}[1]{\left({#1}\right)}
\newcommand{\sz}[1]{\left[{#1}\right]}
\newcommand{\kz}[1]{\left\{{#1}\right\}}
\renewcommand{\sp}{\quad,\quad}
\renewcommand{\v}[1]{\mathbf{#1}}
\newcommand{\m}[1]{\mathrm{#1}}
\renewcommand{\c}[1]{\mathcal{#1}}

\renewcommand{\r}[1]{(\ref{#1})}
\newcommand{\eq}[1]{eq.~(\ref{#1})}
\newcommand{\eqs}[2]{eqs.~(\ref{#1}) and (\ref{#2})}
\newcommand{\eqss}[2]{eqs.~(\ref{#1})--(\ref{#2})}
\newcommand{\Eq}[1]{Eq.~(\ref{#1})}
\newcommand{\Eqs}[2]{Eqs.~(\ref{#1}) and (\ref{#2})}
\newcommand{\Eqss}[2]{Eqs.~(\ref{#1})--(\ref{#2})}

\newcommand{\comm}[2]{\left[{#1}\,,{#2}\right]}
\newcommand{\ket}[1]{\,|{#1}\rangle}
\newcommand{\bra}[1]{\langle{#1}|}
\newcommand{\obs}[1]{\langle{#1}\rangle}
\newcommand{\braket}[2]{\langle{#1}|{#2}\rangle}
\newcommand{\rrho}{\overline{\rho}}
\newcommand{\al}{\hat{\alpha}}
\newcommand{\alp}{\hat{\alpha}^\dagger	}
\newcommand{\hrho}{\hat{\rho}}

\newcommand{\elte}{ELTE, E{\"o}tv{\"o}s Lor{\'a}nd University, H - 1117 Budapest, P{\'a}zm{\'a}ny P. s. 1/A, Hungary}
\newcommand{\kfki}{MTA KFKI RMKI, H-1525 Budapest 114, POBox 49, Hungary}

\title{Multi-boson correlations using wave-packets II.}
\author{M.~I.~Nagy\inst{1}\fnmsep\thanks{\email{nagymarci@rmki.kfki.hu}}
   \and T.~Cs{\"o}rg\H{o}\inst{2}\fnmsep\thanks{\email{csorgo@sunserv.kfki.hu}}}
\institute{\elte\and\kfki}
\abstract{
We investigate the analytically solvable pion-laser model, and its generalization to arbitrary multiplicity distributions.
 Although this kind of extension of the model is possible, the pion laser model in its original form is unique: 
it is the only model in its class that posesses an analytic solution.
}
\maketitle
{\it
This work is dedicated to the memory of the late J.  Zim\'anyi 
and corresponds to a follow-up work  on one of his
favourite topics: the analytic solution of the pion laser model.
}
\section{Introduction}

Correlation between different identical particles from a thermalized source is due to the
bosonic or fermionic nature of them. The effect was discovered by R. Hanbury Brown and
Twiss in 1954, they pointed out that the observed correlation of photons carry information
on the angular diameter of the emitting distant star~\cite{HBT}. In experimental high-energy physics,
the phenomenon is known as GGLP effect~\cite{GGLP}, one commonly uses the term HBT-effect
for the fact, that boson correlations appear and the shape of the correlation function is
related to the geometry of the source. Since its discovery the HBT effect was proven to be
an important tool in high energy physics for investigating the space-time extent of the
reaction process. 

In heavy ion physics, experiments usually measure charged pion correlations. If one
neglects the final state interaction (FSI) and multi-particle correlation effects,
the description of the correlation is simple: the correlation function is essentially
the Fourier transform of the emission source distribution, referred to as source
function. But in realistic situations neither FSI nor multi-particle correlations are negligible.
Two-particle FSI can be taken into account either by generalized Coulomb corrections or by
using the so-called imaging
method~\cite{Brown:1997ku}. Multi-particle correlation means that the final
state outgoing wave-function needs to be symmetrized in all variables. This
obviously requires $n!$ operation, where $n$ is the number of particles. If
the phase-space density of produced pions increases, then the effects arising
from multi-particle symmetrization become significant, so at first sight this
is a non-polynomially (NP) hard  problem. Note that the computing time of NP
hard problems  of the order "$n$" increases faster than any power of $n$.
Hence these problems for large $n$ are essentially unsolvable numerically.

But in fact, there are such models in which one can overcome this difficulty and
calculate the correlation functions using fully symmetrized wave-functions.
The first success in this direction was achieved by S.~Pratt~\cite{Pratt:1993uy},
who showed that in a special case of source functions and factorization, the
NP-hard problem reduces to a set of recurrence relations, called ring algebra,
detailed first in ref.~\cite{Chao:1994fq}.
The next step were refs.~\cite{Zimanyi:1997ur,Csorgo:1997us}, where the model was
generalized to wave-packet states instead of plane-wave final states, and the 
recurrence relations were solved. The model had only one flaw: the multiplicity
distribution of it is very special, and in some sense, unphysical. For example,
coherent behaviour corresponds to Poisson multiplicity distribution
in case of optical the optical lasers, while fully chaotic, rare gas limit
corresponds to the Poisson multiplicity distribution in case of the pion laser model.
In the present work we investigated if the original model of
refs.~\cite{Zimanyi:1997ur,Csorgo:1997us,Pratt:1993uy,Chao:1994fq} can be modified
in such a way, that the Poisson multiplicity distribution will characterize the
pion condensate, and not the chaotic state.

In sections \ref{s:def}, \ref{s:obs}, \ref{s:sol} and \ref{s:incl} we
recapitulate and slightly re-formulate earlier results
of refs.~\cite{Zimanyi:1997ur,Csorgo:1997us,Pratt:1993uy,Chao:1994fq} to prepare
their generalization, utilizing the idea of Poisson transformation. Section \ref{s:p_n}
describes the outcome of this investigation and Section \ref{s:summary} summarizes the results.

\section{Definition of the model}\label{s:def}

The investigated pion-laser model describes a multi-particle system containing
arbitrary number of bosonic wave-packet states. The notations $\xi$, $\chi$ and
$\sigma$ stand for the center in coordinate space, the center in momentum space
and the width in momentum space of a given wave packet, respectively, and
$\alpha:=\z{\xi,\chi,\sigma}$ is an abbreviated notation for these prarmeters
(for example, integrating over $\alpha$ means integration over all variables).
The quantities $\xi$ and $\chi$ are three-vectors. Such a one-particle state
$\ket{\alpha}$ characterized by the parameters $\alpha=\z{\xi,\chi,\sigma}$
is created from the $\ket{0}$ vacuum state by $\alp$, the following wave-packet
creation operator:
\be
\alp=\int\frac{\m{d}^3 x}
{\z{\sigma\sqrt{\pi}}^{3/2}}\exp\kz{-\rec{2\sigma^2}\z{p-\chi}^2-i\xi\z{p-\chi}}\hat{a}^\dagger(p) ,
\ee
where $\hat{a}^\dagger(p)$ is the usual pion creation operator. The normalization
of the states $\ket{\alpha}$ is $\braket{\alpha}{\alpha}=1$, while states with
different parameters have overlap, which we denote by
$\braket{\alpha_i}{\alpha_j}:=\gamma_{ij}$. With this, the properly normalized $n$-particle states are 
\bl{e:n-state}
\ket{\alpha_1\dots\alpha_n}=\z{\sum_{\sigma}\prod_{i=1}^n\gamma_{i\sigma_i}}^{-1/2}\prod_{i=1}^n\alp_i\ket{0} ,
\ee
where the summation runs over all the possible $\sigma$ permutations of the $n$ indices, and $\sigma_i$ means the 
image of the $i$ index by the permutation $\sigma$. We note the expression of $\gamma_{ij}$ (if $\sigma_i=\sigma_j=\sigma$): 
\be
\gamma_{ij} = \exp\kz{-\frac{\sigma^2\z{\xi_i-\xi_j}^2}{4}-\frac{\z{\chi_i-\chi_j}^2}{4\sigma^2}+\frac{i}{2}\z{\chi_i-\chi_j}\z{\xi_i+\xi_j}} .
\ee
The investigated model is defined through its density matrix $\hrho$ as
\be
\hrho=\sum_{n=0}^\infty p_n\hrho_n \sp \m{Tr}\hrho=1 
\ee
with $p_n$ being the multiplicity distribution, and $\hrho_n$ the density matrices of events with fixed number
of particles. The normalization is as usual: $\m{Tr}\hrho_n=1$ for any $n$, so $\sum_{n=0}^\infty p_n=1$.
The $\hrho_n$ matrices are built up of the multi-particle states defined above:
\bl{e:n-densmatrix}
\hrho_n=\int \m{d}\alpha_1\dots\m{d}\alpha_n\rho\z{\alpha_1\dots\alpha_n}\ket{\alpha_1\dots\alpha_n}\bra{\alpha_1\dots\alpha_n} ,
\ee
where $\rho\z{\alpha_1\dots\alpha_n}$ are density functions. There is a special type of them, in what case the model is analytically solvable~\cite{Zimanyi:1997ur}, namely, if the normalization
factor in \Eq{e:n-state} is cancelled:
\bl{e:n-density}
\rho_n\z{\alpha_1\dots\alpha_n}=\rec{\c{N}(n)}\z{\sum_{\sigma}\prod_{j=1}^n\gamma_{j\sigma_j}}\prod_{i=1}^n\rho_1(\alpha) .
\ee
Finally, the one-particle densities $\rho_1\z{\alpha}$ are assumed to be Gaussian:
\bl{e:1-density}
\rho_1(\alpha_1)=\delta\z{\sigma_1-\sigma}\times\rec{\z{2\pi R^2}^{3/2}}\exp\kz{-\frac{\xi_1^2}{2R^2}}
\times\rec{\z{2\pi mT}^{3/2}}\exp\kz{-\frac{\chi_1^2}{2mT}} 
\ee
with $T$ being the temperature and $R$ the radius.
The normalization factor in \Eq{e:n-density} is
\be
\c{N}(n)=\sum_{\sigma}\prod_{i=1}^n\int\m{d}\alpha_i\,\rho_1(\alpha_i)\gamma_{i\sigma_i} .
\ee
The definition of the model is completed with the specification of the multiplicity
distribution $p_n$, which step is postponed to Section \ref{s:p_n}.

\section{Observables}\label{s:obs}

We want to calculate inclusive $n$-particle distributions as well as
exclusive ones (considering events only with fixed $m$ number of particles). The definitions of the exclusive and inclusive quantities are, respectively 
\bea 
N_i^{(m)}\z{k_1,\dots,k_n} &=& 
\m{Tr}\kz{\hrho_m\hat{a}^\dagger(k_1)\hat{a}(k_1)\dots\hat{a}^\dagger(k_n)\hat{a}(k_n)} , \label{e:excN} \\
N_i\z{k_1,\dots,k_n}       &=& 
\m{Tr}\kz{\hrho\hat{a}^\dagger(k_1)\hat{a}(k_1)\dots\hat{a}^\dagger(k_n)\hat{a}(k_n)}   =
\sum_{m=i}^{\infty}p_mN_i^{(m)}\z{k_1,\dots,k_n} . \label{e:inclexcl}
\eea
The normalizations are
\bea
\int\m{d}^3k                  & N_i^{(m)}\z{k,k_1,\dots,k_{i-1}} &= \z{m-i+1}N_{i-1}^{(m)}\z{k_1,\dots,k_{i-1}} , \\
\int\m{d}^3k_1\dots\m{d}^3k_i & N_i^{(m)}\z{k_1,\dots,k_i}       &= \frac{m!}{(m-i)!} , \\
\int\m{d}^3k_1\dots\m{d}^3k_i & N_i\z{k_1,\dots,k_i}             &= \obs{n(n-1)\dots(n-i+1)} , 
\eea
where $\obs{\z{n}\dots\z{n-i+1}}$ is the $i$-th factorial moment of the multiplicity distribution.
Because of the special choice of the density functions in
\eq{e:n-density} in the present model it is possible to write down a relatively simple expression for the
exclusive distributions with the auxiliary function
\be
\rrho(p,q)=\int\m{d}\alpha\,\braket{\alpha}{p}\rho_1(\alpha)\braket{q}{\alpha} .
\ee
Performing some combinatorics one finds that
\be
N_n^{(n)}\z{k_1,\dots,k_n}=\frac{n!}{\c{N}(n)}\sum_{\sigma}\prod_{i=1}^n\rrho(k_i,k_{\sigma_i}) .
\ee
From this one obtains the lower order exclusive functions as
\be
N_m^{(n)}\z{k_1,\dots,k_n}=\frac{n!}{(n-m)!}\rec{\c{N}(n)}\times \int\m{d}^3k_{m+1}\dots\m{d}^3k_n\sum_{\sigma}\prod_{i=1}^n\rrho(k_i,k_{\sigma_i}) .
\ee
For sake of clarity we write down the form of the two lowest order functions:
\bea
N_1^{(n)}\z{k_1}    &=& 
  \frac{n}{\c{N}(n)}\int\m{d}^3k_2\dots\m{d}^3k_n\sum_{\sigma}\prod_{i=1}^n\rrho(k_i,k_{\sigma_i})     , \label{e:N_1}\\
N_2^{(n)}\z{k_1,k_2}&=&
  \frac{n(n-1)}{\c{N}(n)}\int\m{d}^3k_3\dots\m{d}^3k_n\sum_{\sigma}\prod_{i=1}^n\rrho(k_i,k_{\sigma_i})  \label{e:N_2} .
\eea
The calculation of these $n!$-termed integrals is usually a NP-hard problem.
However, in the present model a nice analytic solution is possible by means of
a so-called ,,ring algebera''~\cite{Pratt:1993uy}. The generalization to the case of
wave-packets is found in Ref.~\cite{Zimanyi:1997ur}, in the next section
we summarize the results.

\section{Analytic solutions}\label{s:sol}

The integrals in \eqs{e:N_1}{e:N_2} can be evaluated using Wick's theorem: the $\rrho$
functions in a term of te sum over the permutations can be ordered to rings (i.e. orbits
of the given permutation). The evaluation of the integrals then reduces to the evaluation
of this so-called ring algebra. We just quote the results: one introduces the set
of functions $G_i\z{p,q}$ as
\be
G_1(p,q)=\rrho(p,q) \sp G_n(p,q)= \int\m{d}^3p_1 \rrho(p,p_1)G_{n-1}(p_1,q) . \label{e:Gn_recursive}
\ee
These functions appear as ,,ring integrals''. 
Another important quantities are the normalization constants $\c{N}\z{n}$. Following Ref.~\cite{Zimanyi:1997ur} (but slightly deviating from its conventions), we use an alternative notation as $\omega_n=\rec{n!}\c{N}\z{n}$. 
Putting all this together, the one- and two-particle exclusive distribution functions are given by
\bea
N_1^{(n)}(k_1)           &=& \rec{\omega_n}\sum_{i=1}^{n}
                                  G_i(k_1,k_1)\omega_{n-i} ,\label{e:N1n} \\
N_2^{(n)}(k_1,k_2)       &=& \rec{\omega_n}\sum_{i=2}^{n}\sum_{l=1}^{i-1}
                                 \kz{G_l(k_1,k_1)G_{i-l}(k_2,k_2)+G_l(k_1,k_2)G_{i-l}(k_2,k_1)}\omega_{n-i} . 
\eea
The general $N_i^{(n)}$ could be expressed in a much similar way.
The functions $G_n\z{p,q}$ are found to have an analytic form~\cite{Zimanyi:1997ur}.
The key point is the stability property of the Gaussian distribution: the wave-packets as well as the density functions in
\eqs{e:n-densmatrix}{e:n-density} are Gaussians, so the $G_n\z{p,q}$ functions will be also Gaussian, so we
parametrize them as
\be
G_i(p,q)=h_i\exp\kz{-a_i\z{p^2+q^2}+g_ipq} ,
\ee
and then from the recursive definition in \eq{e:Gn_recursive} one can
derive the recurrence relations, which determine the quantities $h_i$, $a_i$ and $g_i$:
\bea
h_{n+1}=h_1h_n\z{\frac{\pi}{a_1+a_n}}^{\frac{3}{2}} &,& h_1=\rec{\z{\pi\sigma_T^2}^{3/2}} , \\
a_{n+1}=a_1-\frac{g_1^2}{4\z{a_1+a_n}}      &,& a_1=\rec{2\sigma_T^2}+\frac{R_{eff}^2}{2} , \\
g_{n+1}=\frac{g_1g_n}{2\z{a_1+a_n}}         &,& g_1=R_{eff}^2 .
\eea
The constants $\sigma_T$ and $R_{eff}$ in the initial values can be considered as effective
values of the source parameters:
\be
\sigma_T^2=\sigma^2+2mT \sp R_{eff}^2=R^2+\frac{mT}{\sigma^2\sigma_T^2} .
\ee
The solution of the recurrence relations is given by the following equations:
\be
a_n = \frac{g_1}{2}\frac{Y_{n+1}}{Y_n}-a_1 \sp
h_n = h_1^n\sz{\z{\frac{2\pi}{g_1}}^n\rec{Y_n}}^{3/2} \sp
g_n = \frac{g_1}{Y_n} , 
\ee
where the auxiliary quantities are
\be
Y_n   = \frac{r_+^n-r_-^n}{r_+-r_-} \sp
r_\pm = \frac{2a_1}{g_1}\pm\sqrt{\z{\frac{2a_1}{g_1}}^2-1} .
\ee
These quantities give the auxiliary $G_i$ functions. 
We do not go into the details of the derivation, just recommend
Ref.~\cite{Zimanyi:1997ur} for a thorough discussion. We should note that the definitions
and equations used in the present paper differ slightly from those of Refs.~\cite{Zimanyi:1997ur,Csorgo:1997us},
as the multiplicity distribution is introduced here in the last possible step.
 
For the exclusive spectra one needs the $\omega_n$ quantities. They obey the following relation, obtained from \eq{e:N1n}
using the normalization condition of $N_1\z{k}$:
\bl{e:wC}
\omega_n=\rec{n}\sum_{i=1}^n\omega_{n-i}iC_i \sp C_n=\int \m{d}k G_n\z{k,k} .
\ee
The solution to this can be obtained from the expansion of the following power series:
\bl{e:omegaC}
\tilde{g}(z):=\sum_{n=0}^{\infty}\omega_nz^n=\exp\z{\sum_{n=1}^{\infty}C_nz^n} .
\ee
We do not go in details now, just mention that in two limiting cases the $\omega_n$ quantities have a simple form:
if $R_{eff}^2\sigma_T^2\to\infty$ (that is, in the rare gas limit) 
\bl{e:omegararelimit}
C_n=\delta_{n,1} \quad\Rightarrow\quad \omega_n=\rec{n!} ,
\ee
while in the opposite, dense gas limiting case (that is, $R_{eff}^2\sigma_T^2\to 0$)
\bl{e:omegadenselimit}
C_n=\rec{n} \quad\Rightarrow\quad \omega_n=1 .
\ee

\section{Inclusive spectra}\label{s:incl}

In a real experimental situation one usually measures the inclusive spectra,
exclusive measurements can be done much harder. However, for the inclusive spectra
ne needs the exclusive ones, and so one needs to specify the multiplicity
distribution, and at this point difficulties occur. 

We could specify any kind of multiplicity distribution, but \eq{e:omegaC}
suggests a way that enables us to continue the calculation analytically. Namely, the role of the $C_n$-s
in \eq{e:wC} is very similar than that of the so called combinants of a multiplicity distribution:
in this sense $\omega_n$ corresponds to the multiplicity distribution itself. So one reasonable
possibility is to choose $p_n$ as 
\bl{e:pn_original}
p_n=\z{\sum_{m=0}^\infty\omega_m\overline{n}_0^m}^{-1}\omega_n\overline{n}_0^n .
\ee 
where $\overline{n}_0$ is some ,,mean'' value. We note two important relations concerning infinite sums:
\bl{e:sums1}
\z{\sum_{k=0}^\infty a_k}\z{\sum_{l=1}^\infty b_l}=\sum_{k'=1}^\infty\z{\sum_{l'=1}^\infty a_{k'-l'}b_{l'}} ,
\ee
\bl{e:sums2}
\z{\sum_{k=0}^\infty a_k}\z{\sum_{l=1}^\infty b_l}\z{\sum_{q=1}^\infty b_q}
=\sum_{k'=2}^\infty\z{\sum_{l'=2}^\infty\sum_{q'=1}^\infty a_{k'-l'}b_{l'-q'}c_{q'}} .
\ee
Using these, the inclusive spectra are found to have a simple form~\cite{Zimanyi:1997ur}.
Introducing the function 
\be
G\z{p,q}=\sum_{n=0}^\infty G_n\z{p,q}\overline{n}_0^n ,
\ee
one has
\bea
N_1\z{k_1}          &=& G\z{k_1,k_1} , \\
N_2\z{k_1,k_2}      &=& G\z{k_1,k_1}G\z{k_2,k_2}+G\z{k_1,k_2}G\z{k_2,k_1} .
\eea
Higher order functions can be obtained in a much similar way. 
It is important to note two specialities of these expressions. First, since $G\z{p,q}$ has
an analytic form, all the measurable inclusive distribution functions can be calculated in
an easy way. Second, evaluating $G\z{p,q}$ at the intercept point $p=q$, we get $G\z{p,p}=1$,
so, for example, if we define the two-particle correlation function as 
\be
C\z{p,q}:=\frac{N_2\z{p,q}}{N_1\z{p}N_1\z{q}} ,
\ee
we see, that it has a physical intercept behavior, $C\z{p,p}=2$. 
Similarly, for the higher-order correlation functions, $C_n(p, ... , p) = n!$.

This multiplicity distribution was considered in Refs.~\cite{Zimanyi:1997ur,Csorgo:1997us}. Our aim was to generalize the
multiplicity distribution and find analytic expressions for the inclusive spectra.  A motivation arises from the limiting
cases of the present multiplicity distribution: we see from \eqs{e:omegararelimit}{e:omegadenselimit} that in the rare gas 
limiting case $p_n$ approaches a Poissonian distribution with $\overline{n}_0$ mean, and in the opposite, dense gas case
we have
\be
p_n=\frac{\obs{n}^n}{\z{\obs{n}+1}^{n+1}} \sp \obs{n}=\frac{\overline{n}_0}{1-\overline{n}} .
\ee
We would be interested in a distribution which has the opposite behavior: in the rare gas limit it should be a thermal, 
negative binomial distribution, and in the dense limit a condensed, laser-like Poissonian distribution. We will see that
this goal can be achieved without difficulties by redefining $p_n$, but the analytic simplicity will be lost.

\section{General $p_n$-s, uniqueness}\label{s:p_n}

We can generalize the definition of $p_n$ in \eq{e:pn_original} to a case of an integral transformation as
\bl{e:pntrans}
p_n=\int_0^\infty\m{d}y\z{\sum_{m=0}^\infty\omega_my^m}^{-1}\omega_ny^n H\z{y} ,
\ee
This is a pretty general multiplicity distribution, since the function $H\z{y}$ is arbitrary. If
$H\z{y}=\delta\z{y-\overline{n}_0}$, then $p_n$ is the original one as in \eq{e:pn_original}.
In the rare gas limiting case the transformation is just
\be
p_n=\int_0^\infty\m{d}y\exp\z{-y}\frac{y^n}{n!}H\z{y} ,
\ee
which is the so-called Poissonian transformation of the $H\z{y}$ function. As an example, let us consider the
function $H\z{y}=\rec{\obs{n}}\exp\z{-\frac{y}{\obs{n}}}$: it yields a geometrical distribution in the 
rare gas limiting case:
\be
H\z{y}=\rec{\obs{n}}\exp\z{-\frac{y}{\obs{n}}} \quad\Rightarrow\quad p_n=\frac{\obs{n}^n}{\z{\obs{n}+1}^{n+1}} .
\ee
Another motivation for the choice of the multiplicity distribution as in \eq{e:pntrans}
lies in the calculation method of the inclusive spectra. Before going into this, let us
mention, that if we allow $H\z{y}$ to depend not only on $y$, but also on the model parameters
($P_{eff}^2$ and $\sigma_T^2$), then practically all type of multiplicity distributions can
be expressed as an integral transform according to \eq{e:pntrans}. So for a given multiplicity distribution
one can find the appropriate $H\z{y}$ function.

We can use \eqs{e:sums1}{e:sums2} and their analogues with more variables to calculate the inclusive spectra.
For example, from \eqs{e:inclexcl}{e:sums1} $N_1\z{k_1}$ is 
\bl{}
N_1 \z{k_1}=\int_0^\infty \m{d}y \frac{H\z{y}}{\sum_{m=0}^\infty\omega_my^m} \sum_{i=1}^\infty y^i\sum_{j=1}^{i}
                                  G_j(k_1,k_1)\omega_{i-j}=\sum_{n=1}^\infty G_n\z{k_1,k_1}f_n , 
\ee
where $f_n=\int_0^\infty H\z{y}y^n$. In the same way, we have for the two-particle inclusive function
\bl{e:general_N2}
N_2\z{k_1,k_2}=\sum_{i=0}^\infty\sum_{j=0}^\infty\kz{G_i\z{k_1,k_1}G_j\z{k_2,k_2}+G_i\z{k_1,k_2}G_j\z{k_2,k_1}}f_{i+j} ,
\ee
and so on for $N_n\z{k_1,\dots,k_n}$.

We can now draw two conclusions
\begin{enumerate}
\item If and only if $f_{n+m}=f_nf_m$, then we can introduce the quantity
$\tilde{G}\z{p,q}=\sum_{i=0}^\infty f_iG_i\z{p,q}$, and in this case the
inclusive quantities have a simple analytic form: it is enough to calculate
$\tilde{G}\z{p,q}$ \emph{once}, and all the inclusive functions are simple combinations of it.
\item The intercept behavior of the correlation functions is physical if and only if $f_{n+m}=f_nf_m$.
That is, for example, the requirement for $C_2\z{p,p}$ to be $2$, is met only if the $f_n$ integrals
possess the mentioned relation. 
\end{enumerate}
But, $f_{n+m}=f_nf_m$ holds only for the case, when $H\z{y}=\delta\z{y-\overline{n}_0}$ with some $\overline{n}_0$ value.
So, the conclusion of the present investigation is a negative statement: the multiplicity distribution can be generalized, 
but the analytic simplicity of the inclusive distribution functions as well as the physical behavior of the correlation functions do not remain for a more general $p_n$ than investigated already in Ref.~\cite{Zimanyi:1997ur}.

\section{Summary}\label{s:summary}

The aim of this analysis was to generalize the pion-laser model. The model was
first described by S.~Pratt~\cite{Pratt:1993uy}, and then solved entirely by
one of us (T. Cs.) together with J. Zim\'anyi~\cite{Zimanyi:1997ur,Csorgo:1997us}.
We have found, that although it is possible to modify the definition of the
multiplicity distribution so that the thermal or chaotic limit corresponds to a negative binomial
multiplicity distribution,  but  if we require the analytic solvability of the model, it
uniquely prescribes that the multiplicity distribiton in the rare gas limit has to be a Poisson. This is the only
case, when the intercept of the correlation function in the rare gas limit has the expected $C_2(k,k) = 2$ value.
In this sense, we have proven that the original form of the pion-laser model is unique.
A possible other path for analytically solvable multi-particle
systems could be developed if one omits some of the assumptions of the present model: 
for example, if one factorizes the density function in another way than in \eq{e:n-density}.
But these generalizations --- or more like another models --- are extremely difficult to
handle.
	
After the publication of the solution of the pion laser in
refs.~\cite{Zimanyi:1997ur,Csorgo:1997us,Pratt:1993uy,Chao:1994fq}, a similar generalization
has been proposed by Heinz, Scotto and Zhang in ref.~\cite{Heinz:2000uf}.
Working with other conventions, they also obtained the general form of the inclusive
spectra as in \eq{e:general_N2}, however, the emphasis of their work was not on the
uniqueness  of the analytic solution, but more on its numerical generalization.
With the present work, we in a sense complete the investigations of refs.~\cite{Zimanyi:1997ur,Csorgo:1997us}.
We find that the  pion-laser model is unique, in the sense that it is the only the analytically
solvable  model of multi-boson symmetrization of wave-packets in the class of the considered models.


\begin{thebibliography}{}

\bibitem{HBT}   
  R. Hanbury Brown and R.Q. Twiss,
  Nature {\bf 178}, 1046 (1956)

\bibitem{GGLP}   
  G.~Goldhaber, S.~Goldhaber, W.~Y.~Lee and A.~Pais,
  Phys.\ Rev.\ {\bf 120}, (1960) 300.

\bibitem{Brown:1997ku}
D.~A. Brown and P. Danielewicz, Phys. Lett. {\bf B398},  252  (1997).

\bibitem{Zimanyi:1997ur}
  J.~Zim\'anyi and T.~Cs\"org\H{o},
  Heavy Ion Phys.\  {\bf 9} (1999) 241
  [arXiv:hep-ph/9705432].

\bibitem{Csorgo:1997us}
  T.~Cs\"org\H{o} and J.~Zim\'anyi,
  Phys.\ Rev.\ Lett.\  {\bf 80} (1998) 916
  [arXiv:hep-ph/9705433].

\bibitem{Pratt:1993uy}
  S.~Pratt,
  Phys.\ Lett.\  B {\bf 301} (1993) 159.

\bibitem{Chao:1994fq}
  W.~Q.~Chao, C.~S.~Gao and Q.~H.~Zhang,
  J.\ Phys.\ G {\bf 21} (1995) 847
  [arXiv:hep-ph/9411415].

\bibitem{Heinz:2000uf}
  U.~W.~Heinz, P.~Scotto and Q.~H.~Zhang,
  Annals Phys.\  {\bf 288} (2001) 325
  [arXiv:hep-ph/0006150].

\end{thebibliography}
\end{document}